# 基于小波深度置信网络的风电爬坡预测方法

唐振浩[1]，孟庆煜[1]，曹生现[1]，李　扬[2]，牟中华[3]，庞晓娅[3]

（1. 东北电力大学自动化工程学院，吉林 132012；2. 东北电力大学电气工程学院，吉林 132012；

3. 国网甘肃省电力公司电力科学研究院，兰州 730070）

**摘　要**：为了提高风电爬坡事件预测的准确性，提出一种基于深度学习的具有特征自适应选择的小波深度置信网络（WDBNAFS）算法。首先，分析风电功率混沌特性。然后，对时间序列数据进行小波分解，设计特征自适应选择算法选取建模数据作为预测模型的输入变量。最后，采用深度置信网络构建风电爬坡事件预测模型，设计基于实际生产数据的实验验证所提出算法的有效性。仿真结果表明，所提出算法预测准确率可达 90%以上。

**关键词**：深度学习；特征选择；预测模型；风电爬坡事件；深度置信网络

**中图分类号**：TM614　　　　　**文献标识码**：A

## 0　引　言

根据中国能源网发布的信息，2018 年，我国风电新增并网容量 2033 万 kW，累计并网 1.84 亿 kW。全年上网电量 3570 亿 kWh，全年风电利用小时数 2103 h，同比增加 153 h。风电在整个电力系统的占比逐年递增。但是，由于风具有随机性和波动性等特性，导致短时间内风电功率大幅度波动，即发生风电爬坡事件[1]。风电爬坡事件严重威胁电网的安全稳定运行，如 2008 年在美国德州发生了一起严重的爬坡事件，从而给德州电力可靠性委员会（Electric Reliability Council of Texas，ERCOT）带来了严重的经济损失[2]。因此，精确的风电爬坡事件预测是保障电网安全运行的重要措施。

风电爬坡预测最常见的方式是基于历史风电功率数据进行预测[3, 4]。使用历史风电数据的这种方法的预测结果可把握风电的长期变化趋势，但对局部特征的表现较差，不利于爬坡特征的提取。为了提高模型的准确度，Pinson 等[5]、欧阳庭辉等[6]将气象预测数据和统计模型相结合，来提高模型预测精度。受已有研究启发，采用历史风电数据与气象数据相结合的混合数据进行风电爬坡事件建模预测研究。在建模算法方面，传统机器学习算法难以提取数据深层信息，导致建模精度受限。深度神经网络（deep belief network，DBN）拥有深度特征提取能力强等优点，在风功率预测[7, 8]方面有良好的应用。为了进一步提高模型预测精度，采用深度置信网络进行建模。

直接使用原始风功率数据进行建模虽然具有较好的可解释性，但数据中的有效信息难以被完全挖掘。因此需对原始数据进行数据分解。在数据分解方面，经验模态分解（empirical mode decomposition，EMD）[9, 10]、集合经验模式分解（Ensemble empirical mode decomposition，EEMD）[11]等都有成功应用。小波分解（wavelet decomposition，WD）具有提取时域和频域信息的能力，因此，在本研究中使用小波分解进行风功率数据的分解。为了进一步提高输入特征的有效性，设计自适应输入特征提取算法，降低模型

复杂度。

综上所述，针对风电爬坡事件难以准确预测的问题，本文提出一种具有特征自适应选择的小波深度置信网络（WDBNAFS）算法。

## 1 爬坡事件分析

### 1.1 风功率混沌特性分析

Lyapunov 指数法是一种常用的定量判断数据是否具有混沌特性的方法。Michael 等[12]利用 Lyapunov 中的 wolf 法来判断数据混沌特性的方法，如果最大 Lyapunov 指数小于 0，则说明序列具有周期性或者为定值；如果最大 Lyapunov 进行计算。从图 1 中可发现，对于不同时间延迟和嵌入空间位数的参数组合，风功率数据对应的 Lyapunov 指数都大于 0。这一实验结果表明，风功率数据具有混沌特性。

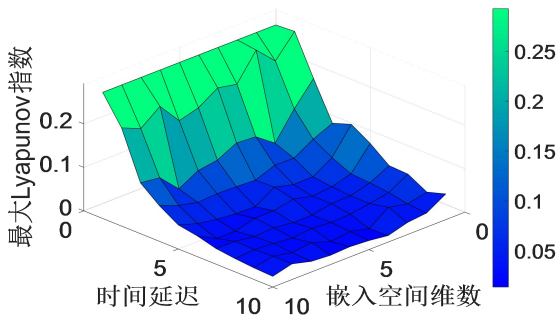

图 1　风功率混沌特性分析图

Fig. 1　Analysis of wind power chaos characteristics

### 1.2 风电爬坡事件分析

目前常见的风电爬坡定义有 4 种[13]，本文所使用的风电爬坡定义是目前应用最为广泛的，计算方法如公式（1）所示：

$$R(t) = \frac{|P(t+\Delta t) - P(t)|}{\Delta t} > H \quad (1)$$

式中，$P(t)$——$t$ 时刻的风电功率，W；$R(t)$——$t$ 时刻的爬坡率，$\Delta t$——时间间隔，min；$H$——爬坡阈值，由于爬坡事件具有方向性，当 $P(t+\Delta t) - P(t) > 0$ 时，该爬坡事件为上行爬坡事件，当 $P(t+\Delta t) - P(t) < 0$ 时，该爬坡事件为下行爬坡事件。

此爬坡定义的 $\Delta t$ 与 $H$ 都是由人工选定，在文献[14]中给出了 $\Delta t$ 的参考值。结合本研究中数据的实际情况，选取 $\Delta t$ 为 30 min，$H$ 取 16。

## 2 风电爬坡预测模型建立

为了建立准确的风电爬坡预测模型，本研究提出基于特征自适应选择的 WDBNAFS 算法。该算法包含特征自适应选择、数据小波分解、DBN 建模 3 个环节。

### 2.1 建模数据重构

本文以某风电场的数据采集与监视控制系统（supervisory control and data acquisition，SCADA）服务器中提取原始数据。首先对原始数据进行相关性分析，结果表明风功率、风速、风向三者之间相关性很大，由于将 3 种特征输入到模型当中进行训练会出现过拟合的状况导致预测结果不佳，而风电爬坡事件是由风功率的突变所产生，并且考虑到温度的变化会对风功率产生影响，所以本文使用历史风电功率及温度作为特征向量。本文的训练数据集分为输入和输出，输入包含风电功率和温度数据，输出是由式（1）所确定的风电爬坡事件。数据集具体结构如表 1 所示。表中 $W(t)$ 为 $t$ 时刻的风机发电功率，$T(t)$ 为 $t$ 时刻的室外温度。其中，功率与温度的时间间隔为 10 min；0 表示未发生爬坡事件，1 表示发生正爬坡事件，-1 表示发生负爬坡事件。由于输入特征的变量量纲差距较大，对所有数据采用 Min-max 方法进行归一化处理。

表 1　重构后数据描述

Table 1　Description of reconstructed data

| 输入 | | 输出 |
|---|---|---|
| 风电功率 $W$/W | 温度 $T$/℃ | |
| $W(t)\ W(t-1)\ \cdots\ W(t-15)$ | $T(t)\ T(t-1)\ \cdots\ T(t-15)$ | 爬坡事件 |

| | | | | | | | |
|---|---|---|---|---|---|---|---|
| 1673.7 | 1267.1 | | 1940.5 | 13.5 | 13.7 | 14.1 | 0 |
| 878.2 | 1673.7 | | 2027.6 | 13.4 | 13.5 | 13.9 | -1 |
| 1168.1 | 878.26 | ⋯ | 1840.8 | 13.3 | 13.4 | ⋯ 13.9 | -1 |
| 1368.1 | 1168.1 | | 1946.8 | 13.3 | 13.3 | 13.9 | 0 |
| 1507.6 | 1368.1 | | 1982.8 | 13 | 13.3 | 13.4 | 1 |

## 2.2 特征自适应选择

在建立预测模型之前，首先要确定模型的输入。结合参数优选思想，设计特征自适应选择法。算法流程图如图2所示。

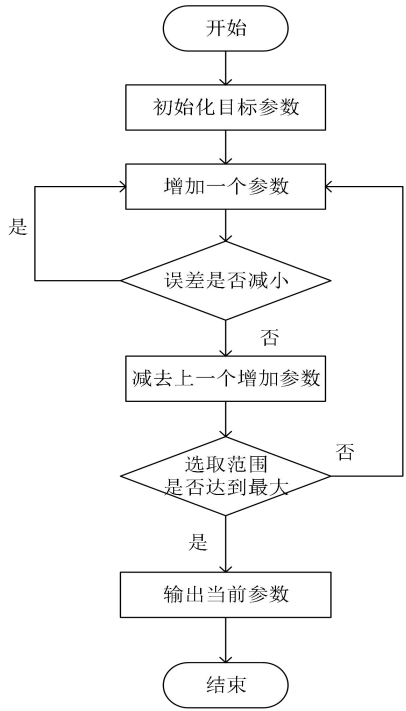

图 2 特征自适应选取算法流程图

Fig. 2 Flow chart of adaptive feature selection algorithm

具体步骤如下：

步骤1：初始化特征子集为空，添加与目标类相关性最大（根据Pearson相关系数确定）的第一个特征；

步骤2：搜索下一个最能降低分类错误率的候选特征：遍历所有不在特征子集里面的输入参数，随机选取一个输入变量，并用新加入变量后的特征子集进行预测建模，若误差减小，则保留所增加参数；若误差增大，不保留该参数；

步骤3：判断是否达到终止条件，如果达到，则停止算法，输出特征子集信息，如果没有，则重复执行步骤2；

步骤3中所描述的终止条件有2个，满足其一即可终止特征选择过程。第1个条件：增加输入参数后，模型预测精度没有提高；第2个条件：所有参数均已加入特征子集。

## 2.3 小波分解

在风电功率的原始时间序列数据当中包含低频信号和高频信号。为了提高原始数据中包含信息在建模过程中的利用率，将原始信号采用小波分解的方法分解为低频信号和高频信号得到，结果如图3所示。图中$S$为原始信号波型，$D_1$分解后的低频信号波，$A_1$、$A_2$、$A_3$为分解后的高频信号波。本研究中，对已有数据的分析结果显示，将原始数据分解到第3层后，继续分解得到的高频信号的信息熵已很小，因此小波分解只进行三层分解，本文对前7时刻（包括当前时刻）的时序历史风功率序列分解为一层低频信号和三层高频信号，得到32（4×8）维数据作为最终的建模输入。

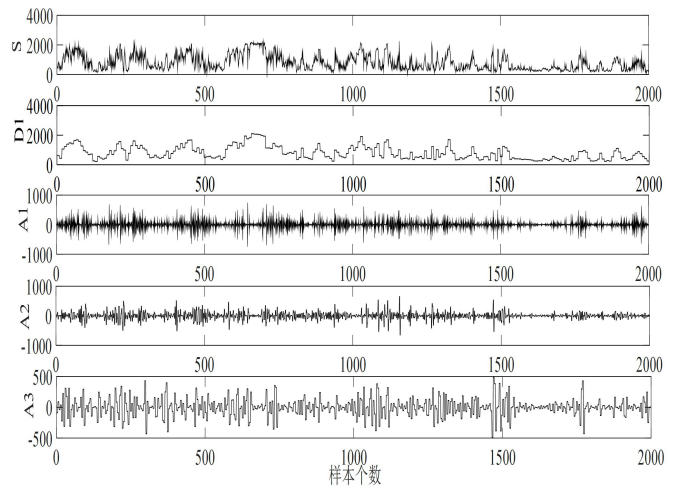

图 3 小波分解结果图

Fig. 3 Results of wavelet decomposition

## 2.4 DBN模型结构

所提出算法中采用的DBN隐含层为2层，每层的单元数为70，其结构如图4所示。DBN由输入层、输出层和隐含层所构成。图4中输入

层和第 1 个隐含层构成 1 个受限玻尔兹曼机（restrict Bolztmann machine，RBM），第一隐含层和第二隐含层组成第 2 个受限玻尔兹曼机。$v$ 代表可见变量，$h$ 代表隐含变量，$w_1$ 表示 RBM1 的权重值，$w_2$ 代表 RBM2 的权重值。经过两层 RBM 学习过程，可以将原始数据中的深层信息提取出来。最后，经过第二隐含层和输出层构成的 BPNN 网络的有监督学习，形成最终的预测网络。

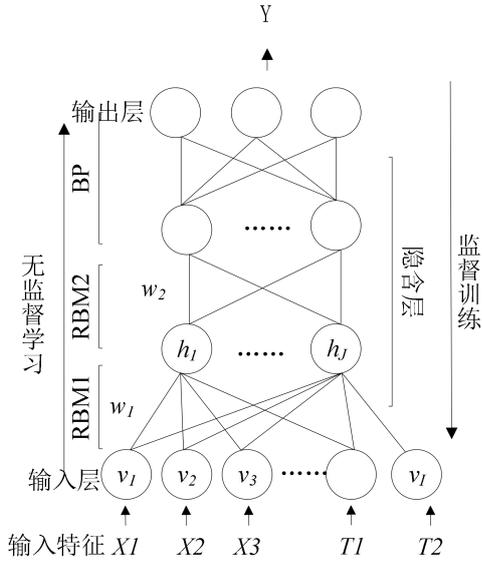

图 4　DBN 结构图
Fig. 4　Structure of DBN

DBN 的训练过程中，是由每一层的 RBM 对数据进行无监督学习，最大限度地拟合输入数据和输出数据。对于给定的（$v$，$h$）其输入层与隐含层的能量函数为：

$$E(v,h|\theta)=-\sum_{i=1}^{I}a_iv_i-\sum_{j=1}^{J}b_jh_j-\sum_{i=1}^{I}\sum_{j=1}^{J}v_iw_ih_j \quad (2)$$

式中，$\theta$——RBM 的参数，$\theta=(w_{ij},a_i,b_j)$；$w_{ij}$——第 $i$ 个输入层节点与第 $j$ 个隐含层节点 $h_j$ 连接的权重值，$a_i$——第 $i$ 个可见层节点的偏置；$b_j$——第 $j$ 个隐含层节点的偏置。可见层和隐藏层的联合概率分布如式（3）所示：

$$P(v,h|\theta)=\frac{1}{z(\theta)}\exp(-E(v,h|\theta)) \quad (3)$$

$$Z=\sum_v\sum_h e^{-E(v,h)} \quad (4)$$

式中，$Z$——归一化函数，由于 DBN 只能确定输出层的状态，而 RBM 的结构原因，在输出层状态确定后隐含层中的各节点的激活状态是相互独立的，所以第 $j$ 个隐含层节点的激活概率为：

$$P(h_j=1|v,\theta)=\sigma(b_j+\sum_{i=1}^{I}v_iw_{ji}) \quad (5)$$

$$\sigma=\frac{1}{1+\exp(-x)} \quad (6)$$

式中，$\sigma$——sigmoid 激活函数，同理第 $i$ 个可见层节点的激活概率为：

$$P(v_i=1|v,\theta)=\sigma(a_i+\sum_{j=1}^{J}h_iw_{ji}) \quad (7)$$

参数 $w_{ij}$ 可按照式（8）进行更新：

$$w_{ij}=\beta w_{ij}+\eta(\langle v_ih_j\rangle_{data}-\langle v_ih_j\rangle_{model}) \quad (8)$$

式中，$\beta$——动量；$\eta$——学习率，取 0.08；$\langle v_ih_j\rangle_{data}$——数据分布期望；$\langle v_ih_j\rangle_{model}$——RBM 模型定义期望，根据吉布斯采样确定 $\langle v_ih_j\rangle_{model}$ 为 0.06。同样地，RBM 其他 2 个参数 $a_i$ 和 $b_j$ 也可通过类似的方式进行更新调整。

当输入数据进入到输入层（其中 $X$ 为小波分解后的风电功率，$T$ 为气象数据），RBM 开始进行无监督学习，根据式（2）~式（4）使第一个隐含层最大的学习并提取原始数据的特征，将训练好的 RBM 输出结果作为下一个 RBM 的输入，按照式（5）~式（7）逐层训练 RBM 直至结束，使用 BPNN 进行有监督学习来优化整个网络参数。在最后的标签层添加标准，按照是否发生爬坡、爬坡事件的类型而直接输出结果 $Y$。

### 2.5　WDBNAFS 的建模整体流程

WDBNAFS 用于预测风电爬坡事件的流程主要是将风电数据通过小波分解，把原始信号分解成 4 个子信号后输入到 DBN 中，以建立预测

模型。图 5 为 WDBNAFS 的建模流程图，其具体步骤如下：步骤 1：数据预处理，将提取的风电数据根据 Min-max 方法进行归一化处理，随后根据式（1）确定风电爬坡事件发生的时间和类型，形成初始的建模数据集；

步骤 2：对 WDBNAFS 网络进行初始化，设置最大迭代次数 $K$=500，当前迭代次数 $k=1$；

步骤 3：将原始数据导入预测模型当中，根据特征自适应选择进行最优预测步长的选取（详见 2.2 节）；

步骤 4：将步骤 3 选取的 7 时刻的风功率序列进行小波分解（详见 2.3 节），获取原始风功率信号的高频和低频特征，并与前 7 时刻的温度数据相结合作为预测模型最终的输入；

步骤 5：将训练样本数据集的 40 维（分解后风功率 32 维+8 维温度数据）特征向量作为输入到第一层 RBM 中，通过式（5）~式（7）使 RBM 逐层进行预训练，根据公式（8）更新参数；

步骤 6：将标注的样本数据集的 40 维特征作为预训练后 DBN 模型的输入，按照对应的标签作为输出。通过 BP 神经网络自顶而下对模型进行有监督学习，直到迭代次数达到设定值为止；

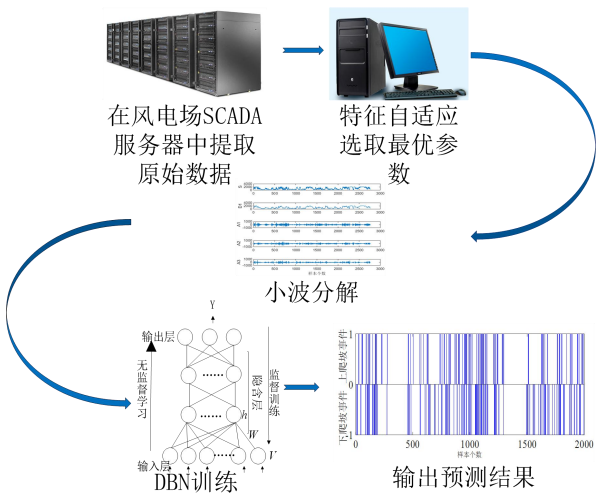

图 5　WDBNAFS 建模流程图

Fig. 5　WDBNAFS modeling flow chart

步骤 7：将输出的数据根据是否发生爬坡事件进行分类，并与实际值情况来进行比对，查看预测结果的精确程度。

## 3 实验分析

### 3.1 数据描述

本文所采用的数据是某风电场一年的实际生产数据，在数据处理阶段，删除数据中的坏点，按照 4 个季度对数据进行划分。将划分好的数据按照 2.2 节方法选取特征参数，而后将原始数据进行小波分解作为预测模型的输入，具体细节如表 2 所示。

表 2　风电数据集特征
Table 2　Characteristics of wind power data sets

| 数据集 | 1 | 2 | 3 | 4 |
|---|---|---|---|---|
| 季度 | 第 1 季度 | 第 2 季度 | 第 3 季度 | 第 4 季度 |
| 采样时间 | 2015-01-01~2015-03-31 | 2015-04-01~2015-06-30 | 2015-07-01~2015-09-30 | 2015-10-01~2015-12-31 |
| 训练集 | 1800 | 1800 | 1800 | 1800 |
| 测试集 | 752 | 797 | 762 | 962 |

### 3.2 评价指标

选取 4 个函数对预测的风电爬坡事件预测评估指标，具体计算公式如式（9）~式（12）所示。$P$ 为正确率（为正确预测出风电事件的概率）、$F$ 为漏报率（发生爬坡事件但未预测出的概率）、$E_1$ 为误报率（未发生爬坡事件但预测发生的概率）、$E_2$ 为错报率（为上、下爬坡预测相反概率）。

$$P = \frac{N_{Pn}}{N_{\text{T}}} \quad (9)$$

$$F = \frac{N_{Fn}}{N_{\text{T}}} \quad (10)$$

$$E_1 = \frac{N_{FP}}{N_{\text{T}}} \quad (11)$$

$$E_2 = \frac{N_{tn}}{N_{\text{T}}} \quad (12)$$

式中，$N_{\text{T}}$——样本总数；$N_{Pn}$——爬坡发生且被预测到的次数；$N_{Fn}$——爬坡发生但未被预测到的次数；$N_{FP}$——爬坡未发生但预测发生的次数，$N_{tn}$——上、下爬坡事件预测相反的次数。

## 3.3 实验结果分析

### 3.3.1 时间窗口选取

通过自适应特征选取对数据进行参数选取，选取后的参数结果如表3所示。表中 $P$ 代表的是风电功率，$T$ 表示温度。从表3可看出，随着对参数进行增加预测误差逐渐减小直至添加参数后误差不再减小。考虑到计算量不宜过大，所以最后温度的时间窗口选取到 $T_{t-7}$，原始风电数据为16维。

表 3 时间窗口选取结果
Table 3　Selection of time window

| 输入变量个数 | 输入变量 | 预测误差/% |
|---|---|---|
| 1 | $P_t$ | 24.00 |
| 2 | $P_t+P_{t-1}$ | 20.10 |
| 3 | $P_t+P_{t-1}+P_{t-2}$ | 19.00 |
| 4 | $P_t+P_{t-1}+P_{t-2}+P_{t-3}$ | 18.80 |
| 5 | $P_t+P_{t-1}+P_{t-2}+P_{t-3}+P_{t-4}$ | 16.80 |
| 6 | $P_t+P_{t-1}+P_{t-2}+P_{t-3}+P_{t-4}+P_{t-5}$ | 15.90 |
| ... | ... | ... |
| 16 | $P_t+P_{t-1}+P_{t-2}+P_{t-3}+P_{t-4}+P_{t-5}+P_{t-6}+P_{t-7}+T_t+T_{t-1}+T_{t-2}+T_{t-3}+T_{t-4}+T_{t-5}+T_{t-6}+T_{t-7}$ | 12.50 |
| 17 | $P_t+P_{t-1}+P_{t-2}+P_{t-3}+P_{t-4}+P_{t-5}+P_{t-6}+P_{t-7}+T_t+T_{t-1}+T_{t-2}+T_{t-3}+T_{t-4}+T_{t-5}+T_{t-6}+T_{t-7}+T_{t-8}$ | 12.50 |

### 3.3.2 建模预测结果及分析

为了验证本文提出算法在不同数据集中的表现，采用4个季度数据分别进行建模。并对不同策略对预测精度的影响进行验证和分析。

1）小波分解对预测结果的影响

为了验证小波分解对实验的影响，本文使用四季度的数据用 DBN 进行预测，并将所得预测结果行对比。图 6 为原始数据+DBN（Raw data+DBN，以 Rd 代替）与经过小波分解后数据+DBN（Wavelet decomposition data+DBN，以 Wd 代替）进行建模的实验结果对比。图 6 按季度划分，分别为第 1 季度、第 2 季度、第 3 季度、第 4 季度。在第 1 季度的数据中，原始数据与小波分解后得到的数据结果类似，两者的预测正确精度相差无几，但在预测误差方面小波分解后结果都优于原始数据；在第 3 季度中原始数据实验结果的漏报数为 41 组，误报数为 5 组，正确个数为 716 组，预测相反数为 0；经过小波分解后的实验结果漏报数为 32 组，误报数为 7 组，正确个数为 723 组，预测相反数为 0；第 2 季度和第 4 季度对比结果可以发现小波分解后的数据建模结果比原始数据直接进行建模的结果好，预测准确率分别提高了 8.16%与 5.82%。经过对比发现小波，小波分解能提高模型的预测精度。

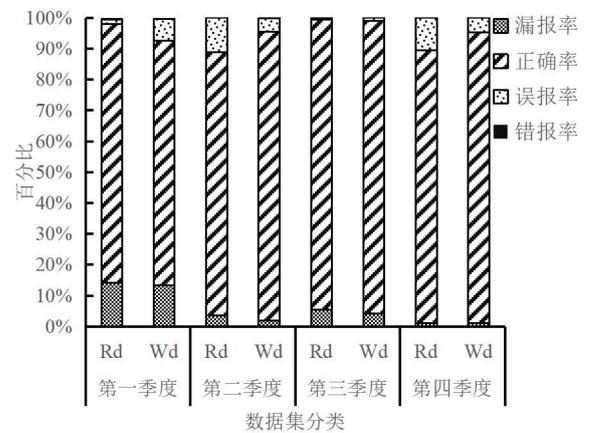

图 6 小波分解前后模型预测结果对比
Fig. 6　Comparison of DBN results before and after WD

2）不同算法预测结果对比分析

为了验证所提出算法的有效性，采用 4 个季度数据进行建模试验。对比不同算法的建模精度。实验结果如图 7 和表 4 所示；图 7 为 DBN 与其他预测算法所得结果的对比图，实验是未经过小波处理的数据所得到的结果。从图 7 可看出，DBN 的预测结果的漏报率低于其他 4 组预测模型；正确率 DBN 的预测结果也比其他模型要高；除 RBF 外，4 组预测模型误报率与错报率的对比结果相差不大。

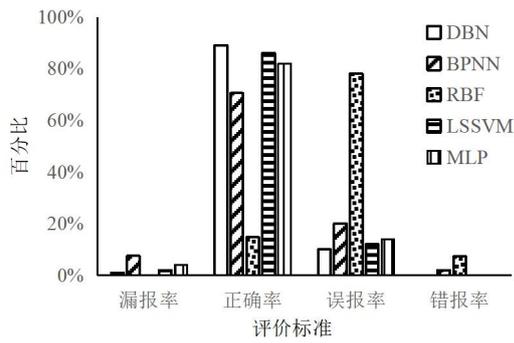

图 7　不同算法实验结果

Fig. 7　Experiment results of different algorithms

表 4　小波分解后不同算法实验结果对比

Table 4　Comparison of experimental results afterwavelet decomposition via different algorithms

| 项目 | 季度 | 漏报率/% | 正确率/% | 误报率/% | 错报率/% |
| --- | --- | --- | --- | --- | --- |
| DBN | 第 1 季度 | 13.43 | 79.12 | 7.04 | 0.41 |
|  | 第 2 季度 | 2.00 | 93.47 | 4.39 | 0.14 |
|  | 第 3 季度 | 4.19 | 94.90 | 0.91 | 0 |
|  | 第 4 季度 | 1.03 | 94.18 | 4.79 | 0 |
| BPNN | 第 1 季度 | 16.62 | 82.71 | 0.40 | 0.27 |
|  | 第 2 季度 | 3.90 | 67 | 28.85 | 0.25 |
|  | 第 3 季度 | 4.08 | 92.25 | 3.67 | 0 |
|  | 第 4 季度 | 8.52 | 79.42 | 11.12 | 0.94 |
| RBF | 第 1 季度 | 0 | 5.72 | 84.31 | 9.97 |
|  | 第 2 季度 | 0.13 | 8.90 | 80.93 | 10.04 |
|  | 第 3 季度 | 0 | 5.51 | 87.41 | 7.08 |
|  | 第 4 季度 | 0 | 8.00 | 83.99 | 8.01 |
| LSSVM | 第 1 季度 | 13.43 | 75.43 | 10.23 | 0.91 |
|  | 第 2 季度 | 3.76 | 81.43 | 14.68 | 0.13 |
|  | 第 3 季度 | 3.80 | 94.88 | 1.32 | 0 |
|  | 第 4 季度 | 2.81 | 85.45 | 11.74 | 0 |
| MLP | 第 1 季度 | 15.84 | 82.84 | 1.06 | 0.26 |
|  | 第 2 季度 | 3.01 | 95.73 | 1.12 | 0.14 |
|  | 第 3 季度 | 4.86 | 87.93 | 7.21 | 0 |
|  | 第 4 季度 | 5.51 | 76.92 | 17.57 | 0 |

从表 4 中可看出，RBF 的预测结果与实际值相差较大，其误报率大。除在第 2 季度数据中 MLP 的预测正确的 763 组比 DBN 预测的 745 组要多之外，DBN 的整体预测结果都要优于其他几种方法。尤其在第 3 季度中，在 762 组试验中 DBN 预测出了 723 组数据，准确率达 94.9%。而在第 4 季度中在发生爬坡事的漏报率为 1.03%，只有 9 组数据漏报。从物理模型和数学统计角度来看，风电爬坡事件是一种小概率事件，其在风功率中占比较低，所以从实验结果来看，DBN+小波的方法对于风电爬坡预测的精度要明显高于其他方法。而 RBF 预测性能明显比其他算法差，表明 RBF 对于时序问题的预测效果差。

由实验结果可知，同一种算法在不同季节进行风电爬坡事件预测的"漏报率"和"误报率"有显著差异，其中第 3 季节的"漏报率"相较于第 2、4 季度较高，而第 3 季度季节"误报率"相较于第 2、4 季度较低。结合数据分析，由于第 3 季度风电功率数据的峰度绝对值相较于第 2、4 季度较小，而峰度的绝对值数值越大则表是其正态分布的陡缓程度与正态分布差异较大，这可能是导致 4 个季度"漏报率"和"误报率"有差异的原因。

为了分析 WDBNAFS 模型对于数据结果分类的精确程度，本文还进行了受试者工作特征（receiver operating characteristic，ROC）曲线分析实验来观察实验结果，结果如图 8 所示。ROC 曲线的横坐标为特异性，是预测错误的概率；纵坐标为敏感性，即预测正确的概率。在 ROC 中判别标准为统计量计算，ROC 曲线的面积值（Area Under the Curve，AUC）在 0.5~1.0 之间，面积越大就表明实验预测的准确性越高。由于本实验为三分类问题，而 ROC 为二分类曲线，所以本文做了 3 组实验即：发生上爬坡事件与未发生爬坡事件（［1，0］）、发生下爬坡事件与未发生爬坡事件（［-1，0］）、发生下爬坡事件与发生上爬坡事件（［-1，1］）。3 组实验两两确定其分类的准确性，图 8 中可看到的 3 条线分别为 AUC=0.5 的分界线、［1，0］、［-1，0］，右下角为局部放大图。粗实线的 AUC 值为 0.992，

表明发生上坡事件与未发生爬坡事件的分类效果比较精确；虚线的 AUC 值为 0.997，表明对于下爬坡事件与未发生爬坡事件的分类更为精确；而实验［-1，1］细实线与图片中左上角的边框重合，其 AUC 值更是达到了 1，表明 DBN 对于上下爬坡事件的分类无误差。根据 ROC 曲线分析表明了 DBN 预测结果非常精确。

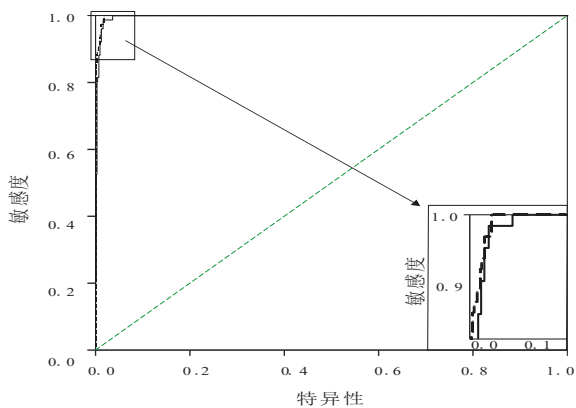

注：粗实线［1，0］为发生上爬坡事件与未发生爬坡事件；虚线［0，-1］为未发生爬坡事件与发生下爬坡事件；与左边框重合细实线［1，-1］为发生上爬坡事件与发生下爬坡事件；对角线为 AUC 面积达到 0.5 的分界线。

图 8 ROC 特性分析图

Fig. 8 ROC analysis chart

## 4 结 论

风电爬坡事件是影响电网稳定性的重要因素。为了实现风电爬坡事件的准确预测，提出一种具有特征自适应选择的小波深度置信网络（WDBNAFS）算法。该算法使用特征自适应选择策略选取输入特征长度，确定模型输入变量。为了获取原始数据的有用信息，对原始信号进行小波分解。根据分解后数据构建建模数据集，并设计深度置信网络构建风电爬坡事件预测方法。使用实际生产数据进行的实验结果表明，与 BPNN、RBF、MLP、LSSVM 等常用预测方法相比，所提出算法的预测结果准确度更高。但是，实验结果同时说明，所提出的预测算法在风电爬坡事件漏报率和误报率方面仍有提升空间。下一步研究将主要进行爬坡事件预测精度的提升以及爬坡事件预测系统的开发及应用推广。

［参考文献］

［1］戚永志，刘玉田.风电高风险爬坡有限度控制［J］.中国电机工程学报，2013， 33（13）： 69—75.

［1］Qi Yongzhi， Liu Yutian.Finite control of high risk wind power ramping［J］.Proceedings of the CSEE，2013， 33（13）： 69—75.

［2］Francis N． Predicting sudden changes in wind power generation［J］.North American WindPower， 2008，5（9）： 58—60.

［3］Taylor J W，Yu K.Using auto-regressive logit models to forecast the exceedance probability for financial risk management ［J］.Journal of the Royal Statistical Society，2016，179（4）：1069—1092.

［4］Taylor J W.Probabilistic forecasting of wind power ramp events using auto regressive logit models ［J］.European Journal of Operational Research，2017 （259）： 703—712.

［5］Pinson P， Madsen H.Ensemble‐based probabilistic forecasting at Horns Rev［J］.Wind Energy， 2009，12（2）： 137—155.

［6］欧阳庭辉，查晓明，秦 亮，等.基于相似性修正的风电功率爬坡事件预测方法［J］.中国电机工程学报，2017，37（02）：572—581.

［6］OuYangTinghui，ZhaXiaoming，Qin Liang，et al.Wind power ramps prediction method based on amendment of similar events［J］.Proceedings of the CSEE，2017，37（02）：572—581.

［7］ZhangYachao，Le Jian，Liao Xiaobing，et.al. A novel combination forecasting model for wind power integrating least square support vector machine, deep belief network, singular spectrum analysis and locality-sensitive hashing［J］.Energy，2019，168：558—572.

［8］WangKejun，QiXiaoxia，LiuHongda，et al. Deep belief network based k-means cluster approach for short-term wind power forecasting［J］.Energy，2018，

165：840—852.

［9］王晓兰，李 辉.基于 EMD 分解的风电场风速和输出功率年度预测［J］.太阳能学报，2011，32（3）：301—306.

［9］ Wang Xiaolan， Li Hui.Annual forecasting of wind speed and power in wind farm based on EMD［J］.Acta Energiae Solaris Sinica，2011，32（3）：301—306.

［10］韩晓娟，田春光，程 成，等. 基于经验模态分解的混合储能系统功率分配方法［J］. 太阳能学报，2014，35（10）：1889—1896.

［10］Han Xiaojuan，TianChunguang，Cheng Cheng，et al.Power allocation method of hybrid energy storage system based on empirical mode decomposition［J］.Acta Energiae Solaris Sinica，2014，35（10）：1889—1896.

［11］Zhang Wenyu，Qu Zhongxin，Zhang Kequan， et al.A combined model based on CEEMDAN and modified flower pollination algorithm for wind speed forecasting［J］.Energy Conversion & Management，2017，136：439—451.

［12］Rosenstein M T，Collins J J，Luca C J D.A practical method for calculating largest Lyapunov exponents from small data sets［M］.Elsevier Science Publishers B.V.1993.

［13］Ferreira C， Gama J， Matias L， et al.A survey on wind power ramp forecasting［J］.Energy & Power Engineering，2011，05（4）：368—372.

［14］Potter C W，Grimit E，Nijssen B. Potential benefits of a dedicated probabilistic rapid ramp event forecast tool［A］. Power Systems Conference and Exposition［C］，Seattle，IEEE，2009：1—5.

# WIND POWER RAMP PREDICTION ALGORITHM BASED ON WAVELET DEEP BELIEF NETWORK

Tang Zhenhao[1]，Meng Qingyu[1]，Cao Shengxian[1]，Li Yang[2]，Mu Zhongha[3]，Pang Xiaoya[3]

（1.School of Automation Engineering， Northeast Electric Power University， Jilin 132012， China；

2.School of Electrical Engineering， Northeast Electric Power University， Jilin 132012， China；

3. State Grid Gansu Electric Power Corporation Electric Power Research Institute， Lanzhou 730070， China）

**Abstract**：The wind power ramp events threaten the power grid safety significantly. To improve the ramp prediction accuracy,a hybrid wavelet deep belief network algorithm with adaptive feature selection (WDBNAFS) is proposed. First,the wind power characteristic is analyzed. Then,a wavelet decomposition is addressed to the time series,and an adaptive feature selection algorithm is proposed to select the inputs of the prediction model. Finally,a deep belief network is employed to predict the wind power ramp event,and the proposed WDBNAFS was testified with the experiments based on the practical data. The simulation results demonstrate that the prediction accuracy of the proposed algorithm is more than 90%.

**Keywords**： deep learning; feature selection; prediction model; wind power ramp events; deep belief network.